\documentclass[namedreferences]{SolarPhysics}
\usepackage[optionalrh]{spr-sola-addons} % For Solar Physics
\usepackage{graphicx}        % For eps figures, newer & more powerfull
\usepackage{color}           % For color text: \color command
\usepackage{url}             % For breaking URLs easily trough lines
\newcommand{\etal}{{\it et al.}}
\newcommand{\degr}{{\hbox{$^\circ$}}}

\newcommand{\asa}{    {\it Astron. Astrophys.}}
\newcommand{\aaps}{   {\it Astron. Astrophys. Suppl.}}

\newcommand{\apj}{    {\it Astrophys. J.}}

\newcommand{\jastp}{  {\it J. Atmos. Sol. Terr. Phys.}}
\newcommand{\jgr}{    {\it J. Geophys. Res.}}
\newcommand{\mnras}{  {\it Mon. Not. Roy. Astron. Soc.}}

\newcommand{\sps}{{\it Solar Phys.}}

\begin{document}
\begin{article}
\begin{opening}
\title{Particle interactions with single or multiple 3D solar reconnecting current sheets}
\author{A.~\surname{Anastasiadis}$^{1}$\sep C.~\surname{Gontikakis}$^{2}$\sep
C.~\surname{Efthymiopoulos}$^{2}$} \runningauthor{Anastasiadis et
al.} \runningtitle{Particle interactions in 3D reconnecting solar
current sheets}
   \institute{$^{1}$ National Observatory of Athens, Institute for Space Applications and Remote
Sensing GR-15236, Penteli, Greece\\
                     email: \url{anastasi@space.noa.gr}\\
              $^{2}$ Academy of Athens, Research Center of Astronomy and Applied Mathematics, \\
Soranou Efessiou 4, GR-11527 Athens, Greece.\\
                     email: \url{cgontik@academyofathens.gr}\\
                     email: \url{cefthim@academyofathens.gr }\\
             }
\begin{abstract}
The acceleration of charged particles (electrons and protons) in
flaring solar active regions is analyzed by numerical experiments.
The acceleration is modelled as a stochastic process taking place
by the interaction of the particles with local magnetic
reconnection sites via multiple steps. Two types of local
reconnecting topologies are studied: the Harris-type and the
X-point. A formula for the maximum kinetic energy gain in a
Harris-type current sheet, found in a previous work of ours, fits
well the numerical data for a single step of the process. A
generalization is then given approximating the kinetic energy gain
through an X-point. In the case of the multiple step process, in
both topologies the particles' kinetic energy distribution is
found to acquire a practically invariant form after a small number
of steps. This tendency is interpreted theoretically. Other
characteristics of the acceleration process are given, such as the
mean acceleration time and the pitch angle distributions of the
particles.
\end{abstract}
\end{opening}
%-------------------------------------------------
\section{Introduction}
     \label{S-Introduction}
The study of the energetic particle acceleration process during
solar flares still remains an open and challenging problem for
solar physics. Solar flares are the manifestation of the energy
release process in the solar corona and atmosphere. It is well
established that during this energy release process the free
magnetic energy is converted, through magnetic reconnection, into
heating, bulk motion of the flaring plasma and particle
acceleration.

In the framework of the theory of magnetic reconnection
process during solar flares, a number of works in the literature
have addressed the problem of particle acceleration. Several
authors studied a magnetic field topology with an induced
homogeneous electric field, with both fields constant in time,
using analytical methods or numerical integration of test
particles.  The magnetic field topology is either an X-point with
or without the presence of a guide field (Bulanov 1980; Deeg,
Borovsky and Duric 1991; Bruhwiler and Zweibel 1992; Moses, Finn
and Ling 1993; Mori, Sakai and Zhao 1998; Browning and Vekstein
2001; Zharkova and Gordovskyy 2005; Hannah and Fletcher 2006) or a
Reconnecting Current Sheet (RCS) (Martens 1988; Martens and Young
1990; Litvinenko 1996; Zharkova and Gordovskyy 2004). The purpose
is to study the final kinetic energy distribution of accelerated
particles, the condition of adiabaticity of their orbits and the
level of charge separation at the ejection points. More realistic
steady X-point topologies were derived using the MHD equations
(Craig and Litvinenko 2002; Heerikhuisen, Litvinenko and Craig
2002).

In order to model the burst effects many authors incorporate
a time dependent electric field in X-points (Hamilton \etal\ 2003;
2005; Petkaki and MacKinnon 1997; 2007) or adopted a numerical
code as Tajima \etal\ (1987). Wood and Neukirch (2005) adopted an
X-point with a nonhomogeneous electric field that was stronger at
the center of the X-point. Finally, Kliem (1995) used O-points and
X-points combinations to study the effect of the fragmentation of
RCS on particles acceleration. On the other hand, several
studies (e.g. Benz \etal\ 1994; Saint-Hilaire and Benz 2002) show
that more than $40 \%$ of the released energy in solar flares is
deposited to high energetic particles, indicating the close
relation between the particle acceleration process and the way
(and amount) of the energy released in these highly energetic
events. Despite this, a very common approach to building
acceleration models (see Miller \etal\ 1997; Anastasiadis 2002 for
reviews) is to consider that the different processes (i.e. energy
release, acceleration, transport and radiation) are decoupled. The
main reason for such a consideration is the very different
temporal and spatial scales on which the different processes
evolve. This difference notwithstanding, it should not be
neglected that all the processes are interwoven in a way rendering
necessary to develop {\it global models} for solar flares, i.e.,
models taking into account the interplay between the individual
processes.

Parker (1998) first proposed that the free magnetic energy could
be released in the solar corona through {\it multiple dissipation
sites}. This assumption implies essentially the fragmentation of
the energy release in both space and time during solar flares.
As several observations, either related to solar flare
parameters (e.g. Dennis 1985; Benz 1985; Crosby, Aschwanden and
Dennis 1993; Crosby \etal\ 1998; Aschwanden \etal\ 1995; 2000) or
to emission processes in the solar corona (e.g. Mercier and
Trottet 1997; Krucker and Benz 1998), can be interpreted on the
basis of such a scenario, several models of particle acceleration
were developed that incorporate concepts from the general theory
of dynamical systems, such as the {\it complexity} of the energy
release process. For example, Anastasiadis and Vlahos (1991,1994)
assumed the presence of multiple shock waves acting on the
particles. Such shocks have the form of small-scale, short-lived
magnetic discontinuities inside a flaring active region. Another
possibility is the acceleration of particles by multiple DC
electric fields (Anastasiadis, Vlahos and Georgoulis 1997;
Anastasiadis \etal\ 2004; Vlahos, Isliker and Lepreti 2004;
Dauphin, Vilmer and Anastasiadis 2007). Such fields are associated
with the variety of dissipation sites inside an active region. A
simulation of the energy release process can then be made by a
cellular automaton (CA), the dynamical substrate of which
incorporates the assumption that the active region evolves
according to the dynamical laws of a system being on the status of
`self-organizing criticality' (see Isliker, Anastasiadis and
Vlahos 2001 for details on CA models for solar flares).

In the solar corona, large-scale and long lasting current sheets,
in which a large number of particles could be accelerated, are
unlikely to be present. Recent MHD simulations show that large
current sheets are not stable for a long time and are quickly
fragmented (Onofri, Isliker and Vlahos 2006). Furthermore, Hughes
\etal\ (2003) have shown that solar flares can be considered as
cascades of reconnecting small-scale magnetic loops inside an
overall simple large magnetic field topology. In other words, the
large-scale magnetic field topology probably determines the
location of the magnetic energy release regions only on the scale
of the active region. In addition, the complex magnetic
environment of an active region together with the turbulent
photospheric motions should be taken into account, since these
external factors drive the system continuously by adding new
stresses to the existing large-scale topologies. In particular,
such a type of driving leads necessarily to the formation of
multiple reconnecting current sheets (RCS) configurations.

In the present work, we consider the above description as the
basis for constructing numerical simulations of the acceleration
of particles (electrons and protons) within multiple 3D current
sheets developing in an active region. Our basic model is a
sequence of encounter events of a population of particles with
local current sheets with physical parameters changing
stochastically. In order to understand the outcome of such a
process, it is important to study first just one step of the
process, namely the interaction of the particle with just one RCS.
This was the basic motive behind two recent papers of ours
(Efthymiopoulos, Gontikakis and Anastasiadis 2005; Gontikakis,
Efthymiopoulos and Anastasiadis 2006) in which we explored the
particle dynamics and acceleration in a single Harris-type RCS
configuration by numerical and analytical means. A basic outcome
of our study is an analytical formula yielding the maximum
possible kinetic energy gain of the particles passing through the
RCS as a function of the particles' initial kinetic energy and of
the physical parameters of the current sheet. Our formula
generalizes previous formulae given by Speiser (1965) and
Litvinenko (1996), containing the  later as asymptotic
limits. In the present paper we further generalize this formula to
an approximate formula valid for single X-point topologies.
Following the assessment of this `one-step' process, we then pass
to modelling the acceleration of particles inside a complex
flaring

active region as a stochastic process. In this we repeat the
methodology used in Gontikakis, Anastasiadis and Efthymiopoulos
(2007) for Harris-type RCSs, and we provide a further theoretical
analysis of the results obtained there in.

In summary, the simulations are done as follows: particles are
initially interacting with a single RCS. This changes their
kinetic energy distribution. After this interaction, the particles
are considered to perform a free flight following the magnetic
field lines until they reach another RCS present in the active
region. This process is in principle repeated ad infinitum.
Nevertheless, one finds that the kinetic energy distribution tends
to acquire a limiting form following a few such steps, after which
no appreciable change is seen in the distribution. This fact was
observed in our previous study and we here provide a theoretical
explanation for it. Furthermore, we expand our study by
considering also multiple encounters of the particles with local
X-point reconnecting magnetic field configurations. In this we
find again the tendency of the kinetic energy distribution to
reach a limiting form. We thus conclude that this phenomenon is
probably generic, i.e., independent of the details of the
reconnection topology.

In the next section (Section~\ref{S-Set}) we outline the basic
characteristics of our numerical set up. In Section~\ref{S-Single}
the interaction of particles with a single RCS is studied,
followed by our results for the multiple encounters
(Section~\ref{S-Multiple}). Section~\ref{S-Summary} summarizes our
results.

\section{The numerical set-up} %%%%%%%%%%%%%%%%%%%%%%%%%%%%%%%%%%%%%%%%
      \label{S-Set}

We assume two types of magnetic field reconnecting topologies. The
first type is the Harris-type model of Litvinenko and Somov (1993)
that represents one local current sheet within an active region.
The magnetic and electric fields inside one sheet of
half-thickness $a$ are given in Equation~(\ref{fields-H}):
\begin{eqnarray}\label{fields-H}
E &= &(0,0,E)  \nonumber \\
B &= &(-\texttt{y}/a,\xi_\perp ,\xi_\parallel )B_0~~~~~\mbox{for }|\texttt{y}|\leq
a.
\end{eqnarray}
The edges of the current sheet are at $\texttt{y}=\pm a$. The magnetic
field is normalized in units of $B_0$, the value of the main
reconnecting component at the edges. The magnetic field has two
components, parallel and perpendicular to the current sheet plane,
called the `longitudinal' ($\xi_\parallel$) and `transverse'
($\xi_\perp$) component respectively. The second type is an
X-point topology defined by the following equations:
\begin{eqnarray}\label{fields-X}
E &= &(0,0,E)  \nonumber \\
B &= &(-\texttt{y}/a,-(\texttt{x}/a)\, \xi_\perp,\xi_\parallel)B_0~~~~~\mbox{for
}|\texttt{y}|\leq a.
\end{eqnarray}
Note that a Harris-type configuration is a simplified model for,
say, the left domain of an X-point configuration far from the
$\texttt{x}=\texttt{y}=0$ point.

In Efthymiopoulos, Gontikakis and Anastasiadis (2005), we studied
the dynamics of the particles in a Harris-type RCS using the
Hamiltonian equations of motions. We found that, using the
problem's symmetry, the initial 3-degrees of freedom Hamiltonian
can be reduced to a two degrees of freedom Hamiltonian  of a form:
\begin{equation}\label{ham2d}
H={1\over 2}p_y^2 +{1\over 2}(c_4+{1\over 2}y^2)^2 +{1\over 2}
(I_2-\xi_{\perp}z + \xi_{\parallel} y)^2 - \epsilon z
\end{equation}
with $\epsilon=E\, m /(a B_0^2 e)$ being the normalized electric
field ($m$ and $e$ are the particle's mass and charge) and
$x=\texttt{x}/a$, $y=\texttt{y}/a$ are the normalized space
coordinates. The time coordinate, implicit in
Equation~(\ref{ham2d}), is scaled with the gyration period. The
canonical momenta are mapped to velocities via $p_y=\dot y$,
$c_4=\dot z - {1 \over 2} y^2$. $I_2$ is an integral of motion
yielding the velocity of the x-component of motion (missing in
Equation~(\ref{ham2d})), namely:
\begin{equation}\label{i2}
I_2=\dot x - \xi_\parallel y + \xi_\perp z
\end{equation}
In addition, we found (Efthymiopoulos, Gontikakis and Anastasiadis
2005; Gontikakis, Efthymiopoulos and Anastasiadis 2006) that the
maximum kinetic energy gain for one particle can be expressed as a
function of the field parameters, the position of injection into
the sheet and the initial energy of the particle with the
relation:
\begin{equation}\label{emax}
E_{max}= {\epsilon \over \xi_\perp^2}\Big( \xi_\perp\,I_2
+\xi_\parallel \xi_\perp y_{out} + \epsilon +\sqrt {2 \xi_\perp
I_2 \epsilon + 2 \xi_\parallel \xi_\perp y_{out} \epsilon +
\epsilon^2 + 2 \xi_\perp^2 E_0 } \Big)
\end{equation}
where $E_0$ is the initial kinetic energy of a particle injected
at $y=y_0$, the value of $I_2$ is set equal to $I_2 = \dot x_0
-\xi_\parallel y_0$, and the exit of the particle is through the
edge $y=y_{out}$.

The study of the particle orbits in a Harris-type current sheet
given by Equation~( \ref{fields-H}) is performed by integrating
numerically the relevant equations of motions using the
Hamiltonian of the form given in Equation~(\ref{ham2d}). For the
case of an X-point geometry we integrate the Newtonian equations
of motion using Equation~(\ref{fields-X}) for the electromagnetic
field. In all our numerical simulations we consider that the RCS
half-thickness $a$ is the unit of length and the inverse
gyrofrequency $\omega_B^{-1}=m/q B_0$ is the unit of time. For a
typical value of the main magnetic field of $B_0=100G$,
$\omega_B^{-1}\simeq 6 \times 10^{-10}$~sec for electrons and
$10^{-6}$~sec for protons. A super-Dreicer electric field $E$ of
100 V/m is used, which corresponds to a normalized  field of
$\epsilon=10^{-5}$ for electrons and of $\epsilon=(m_p/m_e)\times
 10^{-5} \simeq 1.84\times 10^{-2}$ for protons. The
numerical integration is carried out, for both geometries, until
the particle reaches the edges of the reconnection site which is
at at $y=\pm 1$. Finally we consider particle injections at three
different positions along the y-axis, i.e., from the edges ($y=\pm
0.9$) or from the central plane ($y=0$).

\section{Particle interactions by single RCS}
\label{S-Single}
\begin{figure}    %%%%%%%%%%%%%%%%%% FIGURE 1
   \centerline{\includegraphics[width=1.0\textwidth,clip=]{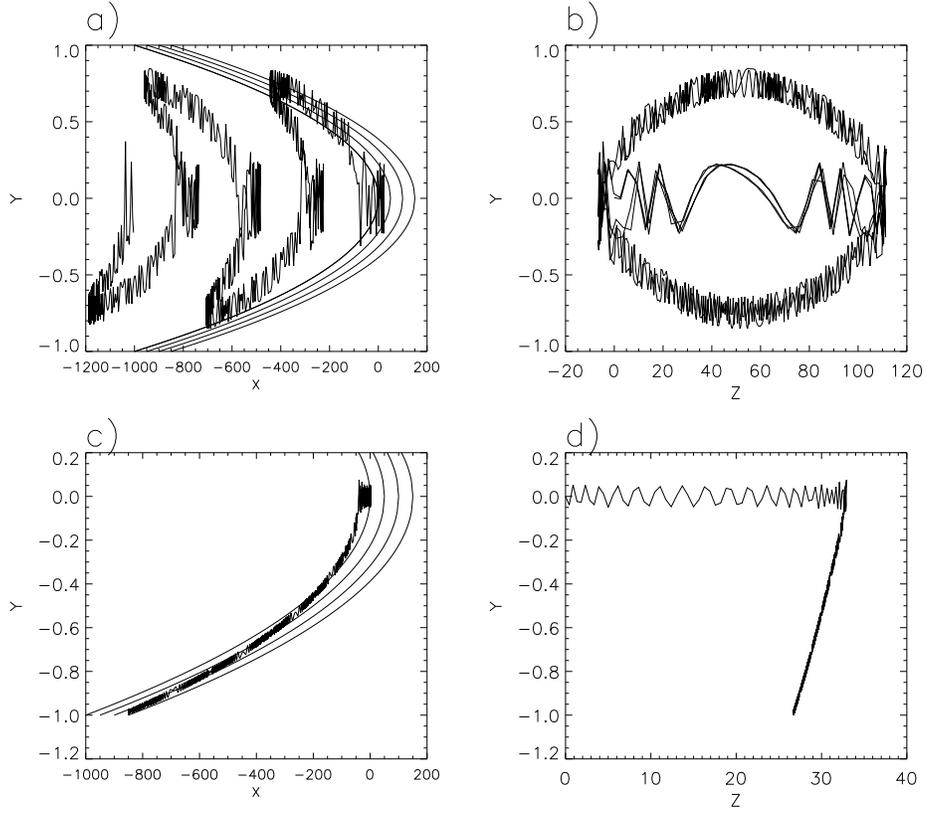}
              }
              \caption{Examples of single electron orbits inside a Harris - type
              reconnecting current sheet with $\xi_{\parallel}=0$ and $\xi_{\perp}=10^{-3}$.
              In panels (a) and (b) the projections of a trapped (mirrored) trajectory are shown.
              In panels (c) and (d) the projections of an escape orbit are presented.
              The continues lines in panels (a) and (c) are visualizations of the magnetic field lines.
                      }
   \label{F-orbits-H}
   \end{figure}
Particles can follow several types of orbits inside reconnecting
magnetic fields. The form of the orbit depends strongly on the
value of the longitudinal component of the magnetic field
$\xi_\parallel$. When $\xi_\parallel$ is less than about 0.1,
electrons can follow chaotic orbits which lead them to escape, or
regular quasi-periodic orbits along which the particle remains
trapped in the sheet. In Figure~\ref{F-orbits-H} two kinds of
electron orbits are shown inside a RCS. In the first case (panels
a,b) the particle never reaches the $y=\pm 1$ edges and performs a
quasi-periodic motion, mirrored by the growth of the magnetic
field strength away from the $y=0$ plane. The second orbit (panels
c,d) is an escaping chaotic orbit. Initially the electron
oscillates in the y-axis with $|y|<0.05$.  Later however, it
leaves the current sheet by moving along a magnetic field line.
For large values of $\xi_\parallel$ most particles follow
adiabatic orbits. In that case the particles nearly follow
adjacent magnetic field lines before escaping from the
accelerating site.

\begin{figure}    %%%%%%%%%%%%%%%%%% FIGURE 2
   \centerline{\includegraphics[width=1.0\textwidth,clip=]{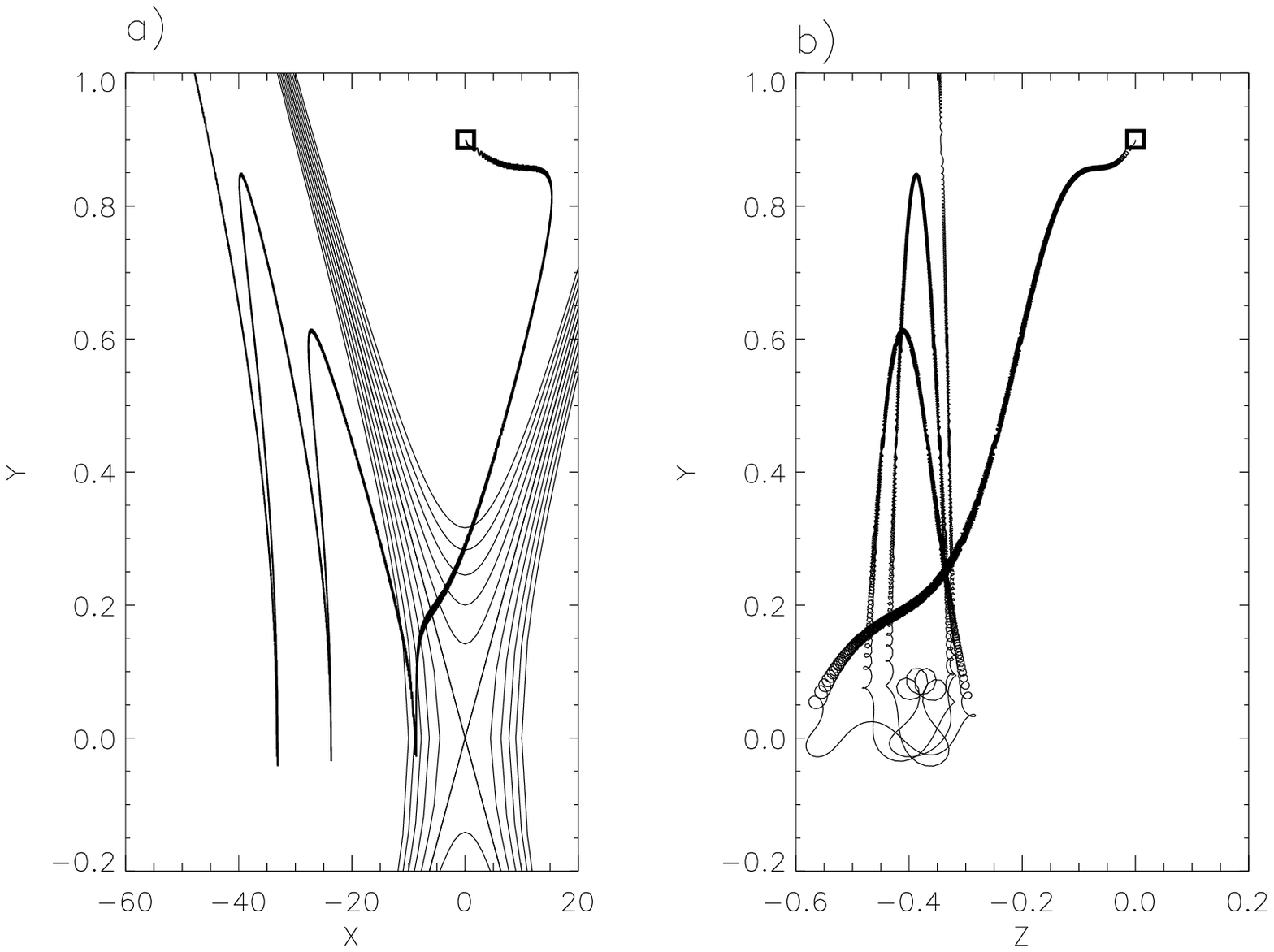}
              }
              \caption{Example of single electron orbit inside
              an X-point configuration, with $\xi_{\parallel}=0$ and $\xi_{\perp}=10^{-3}$.
              The continuous lines in panel (a) are visualizations of the magnetic field lines.
                      }
   \label{F-orbits-X}
\end{figure}

Chaotic and regular (adiabatic) orbits are also found in the case
of acceleration through an X-point configuration. In
Figure~\ref{F-orbits-X} an electron moves chaotically through an
X-point with $\xi_\parallel$=0, $\xi_\perp=10^{-3}$ and
$\epsilon=10^{-5}$. The electron starts its motion at
$(x_0,y_0)=(0,0.9)$ and, then drifts toward the X-point line at
$(x,y)=(0,0)$. There, it performs a bounce motion
 which pushes it to larger values of $y$ until it reaches the $y=1$
edge where it escapes from the current sheet.

We now consider the acceleration of N=30\,000 particles (electrons
or protons) that form initially a thermal distribution at a
typical coronal temperature of $2\times 10^6$~Kelvin. This means
that the velocities of particles are randomly oriented and form a
Maxwellian distribution. Each particle enters from one of
three selected injection points along the y-axis, namely $y=\pm
0.9$ or $y=0$. The initial position of particles on the x-axis is
defined as $|x|<0.5$ for X-points and $x=0$ for Harris-type
current sheets. The field parameters, for both current sheets
configurations, are $\xi_\parallel=1$, $\xi_\perp=10^{-3}$ with
dimensionless electric field parameters $\epsilon=10^{-5}$ for
electrons and $\epsilon=1.84\times 10^{-2}$ for protons.

\begin{figure}    %%%%%%%%%%%%%%%%%% FIGURE 3
   \centerline{\includegraphics[width=1.1\textwidth,clip=]{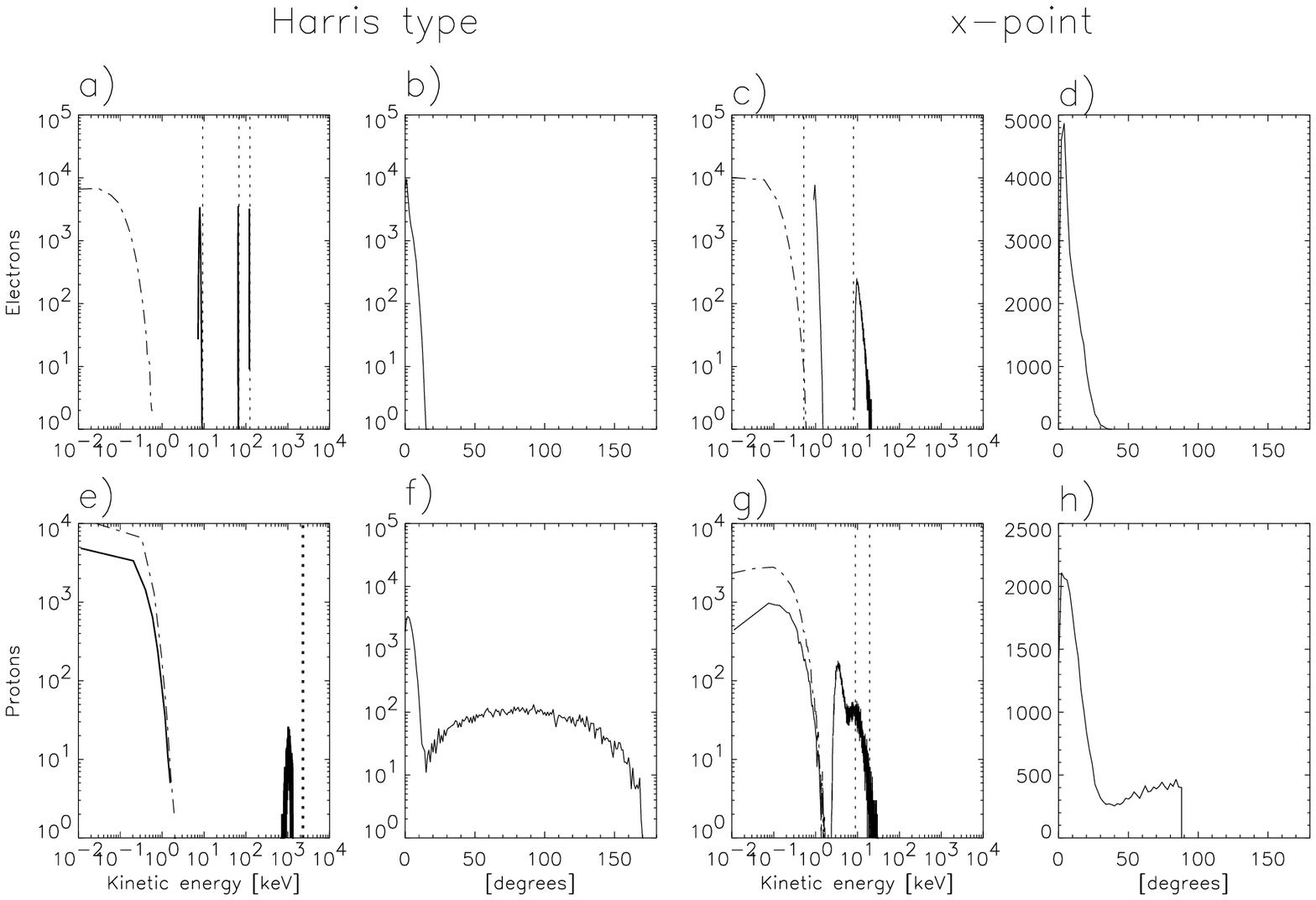}
              }
              \caption{Distributions of kinetic energy and pitch angles for a Harris -
              type (panels a,b, e and f) and X-point  (panels c, d, g and h) RCS
              with $\xi_{\parallel}=1$ and $\xi_\perp=10^{-3}$.
              The first row is for electrons and the second
              row is for protons. In panels (a), (c), (e) and (g) the dashed line
              corresponds to the initially injected kinetic energy distribution
              and the vertical dashed lines to maximum kinetic
              energy derived by the analytical formula given by
              Equation~(\ref{emax})
                      }
   \label{F-distributions}
   \end{figure}

In Figure~\ref{F-distributions}, we present a comparison between
particles accelerated through a Harris-type current sheet and
particles accelerated through an X-point. The presented
characteristics are the final kinetic energy distribution and the
final pitch angle distribution of the particles. In all the
kinetic energy distribution plots, the dotted vertical lines
correspond to the maximum kinetic energy gain computed through
Equation~(\ref{emax}).
Electrons kinetic energy distributions have a short energy range
that is characterized by a number of energy peaks. In the Harris-
type case three energy peaks correspond to the initial injection
positions. In the case of X-points, the distribution presents two
peaks, one for particles injected from the sides ($y=\pm 0.9$) and
one, with higher energy, for injection from the X-point center.
Moreover, for the same parameter values used in both geometries,
particles accelerated with an X-point gain less kinetic energy by
a factor of $\simeq 10$. Because of the fact that the
particles entry points are very localized in the y-axis, the
resulting final kinetic energy distributions are peaked in a short
range of final energies rather than yielding a power-law (like in
Wood and Neukirch (2005); our case best compares with Figure~5 of
that paper, which corresponds to particles accelerated near the
separatrix). On the other hand, our injection of particles at
$y=0$ is representative of particles being accelerated when they
start already inside the current sheet. In fact, if a uniform
distribution of initial conditions is taken in the range $-0.9\leq
y \leq 0.9$ the final kinetic energy distribution fills the gaps
between the peaks of Figure~\ref{F-distributions} a,c.

The analytic formula of Equation~(\ref{emax}) predicts well the
numerical result, especially in the case of the Harris-type
(Gontikakis, Anastasiadis and Efthymiopoulos 2007). In the case of
the acceleration through the X-point on the other hand, the
analytical solution gives a reasonable estimation for the
particles entering from the above edge if we replace the parameter
$\xi_\perp$ with an effective value (see below). As the particle
moves along the x-axis it encounters an increasing value of the
perpendicular component of the magnetic field $B_y=\xi_\perp x$.
Roughly one can replace $\xi_\perp$ with $\xi_\perp <x>$ ($<x>$ is
the average position along the x-axis of a particle orbit) in
Equation~(\ref{emax}) which gives the right estimation of the
kinetic energy gain. The analytic expression is consistent with
the numerical results as long as the particles enter from the
sides of the X-point and is slightly short in energy for electrons
starting at the X-point center (panel c). In fact, as already
mentioned, in the adiabatic limit the projection of the particles'
orbits on the $x-y$ plane follows essentially the projection of
the magnetic field lines on the same plane. The latter is given by
the family of hyperbola:
\begin{equation}\label{mfl}
y^2-\xi_\perp x^2=const~~.
\end{equation}
Particles entering the sheet at $(x_0,y_0)$ and leaving the sheet
at $x_{out},y_{out}$ then satisfy approximately the relation
$|x_{out}|\simeq (1/\sqrt{\xi_\perp})\sqrt{y_{out}^2-y_0^2-\xi_\perp
x_0^2}$. The average transverse magnetic field `felt' by the
particles can then be estimated as:
\begin{equation}\label{xiav}
<\xi_\perp>\approx {\xi_\perp\over 2} \bigg(
(1/\sqrt{\xi_\perp})\sqrt{y_{out}^2-y_0^2-\xi_\perp x_0^2}-x_0 \bigg)~~.
\end{equation}
Equation~(\ref{emax}) can then be used to estimate analytically the
maximum amount of kinetic energy gain for the particles crossing
an X-point, if $<\xi_\perp>$ of Equation~(\ref{xiav}) is substituted in
the place of $\xi_\perp$ in (\ref{emax}).

Protons present a behavior which is quite common in both the
Harris-type and X-point types of accelerators. In particular, a
large fraction of the initially injected protons are not
accelerated, but they simply cross the current sheet practically
without changing their kinetic energy. The final kinetic energy
distribution of these protons is thus still a Maxwellian (see
Figure~\ref{F-distributions} panels e and g). On the other hand,
likewise electrons, the kinetic energy distributions of
accelerated protons present secondary features due also to the
three different positions of injection. Nevertheless, such
features are less apparent because the overall distribution is
broader for protons than for electrons. In the case of X-point,
the kinetic energy distribution of the accelerated protons
presents two components, one associated with an injection from the
sides and the other one for injection from the center of the
current sheet.

In a Harris-type, the final kinetic energy of protons is of order
1~MeV, which is 10 times larger than the energy gain of electrons.
In the case of an X-point, protons and electrons are both seen to
gain considerably less energy, of order 10~keV. As regards exit
pitch angle distributions, both electrons and protons, accelerated
in either X-points or Harris-types current sheets, present similar
narrow distributions (Figure~\ref{F-distributions} b,d,f,h). The
pitch angle distributions of protons exhibit a broad band
corresponding to the population of non accelerated particles, the
velocities of which are randomly oriented both before and after
the interaction with the RCS.

Finally, we studied the time needed for a particle to leave a
current sheet as a function of the longitudinal magnetic field
component $\xi_\parallel$. The time needed for an electron to
leave the current sheet is  of the order or a few micro seconds.
For protons the time is of some milliseconds (Gontikakis,
Anastasiadis and Efthymiopoulos 2007). Note that such time scales
are still out of reach of the time resolution of present
instruments thus little can be said as regards comparison of such
figures with observations.
\begin{figure}    %%%%%%%%%%%%%%%%%% FIGURE 4
   \centerline{\includegraphics[width=1.0\textwidth,clip=]{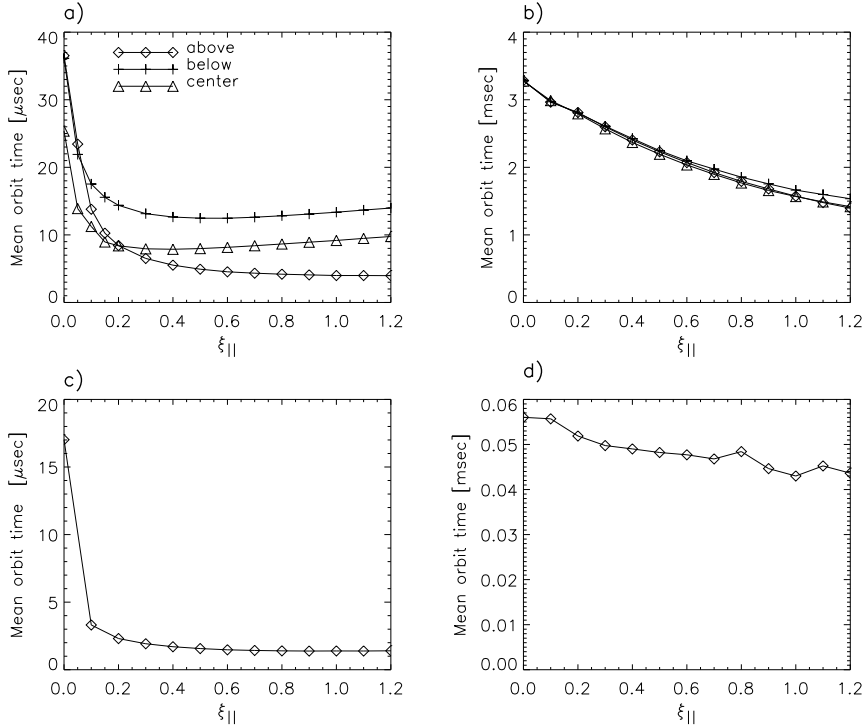}
              }
              \caption{Time spend by accelerated particles inside
              ($|y|<1$) a current sheet depending on
              the value of the $\xi_\parallel$ with $\xi_\perp=10^{-3}$.
              The first row
              concerns acceleration through a Harris-type and the
              second row, acceleration through a X-point. In panels
              (a) and (c) the particles are electrons and in
              panels (b) and (d) protons. At the first row,
              diamonds shows injections from above, crosses
              injections from below and triangles from the center
              of the current sheet. Particles are injected from
              the sides of X-points. The electric field is of $10^{-5}$ and
              $1.84\times 10^{-2}$ for electrons
              and protons respectively. }
   \label{F-meantime}
   \end{figure}
In order to compute the mean time needed for the particles to
reach the edges of current sheets (at $y=\pm 1$), we followed the
orbits of 1000 protons and electrons through a Harris-type and an
X-point current sheet for values of the $\xi_\parallel$ in the
range (0 to 1.2). In Figure~\ref{F-meantime} we present the mean time
as a function of the value of $\xi_\parallel$ for electrons (panel
a) and protons (panel b). For protons, the time inside the current
sheet is a decreasing function of $\xi_\parallel$. For electrons,
the acceleration time decreases sharply when $\xi_\parallel$
changes from 0 to 0.1. This function also depends on the injection
point. Electrons entering the current sheet from above have
acceleration times slowly decreasing with $\xi_\parallel$. On the
other hand, electrons initiated from the center or entering from
below ($y=-0.9$) show a small rise of the acceleration time as a
function of $\xi_\parallel$. In the case of electrons, when
$y_0 \not = 0.9$ the  acceleration time grows asymptotically, for
large $\xi_\parallel$, as a function of $\xi_\parallel$. This
result is in agreement with the asymptotic analysis of Litvinenko
(1996)in the case of strongly magnetized particles.

\section{Particle interactions by multiple RCS}
\label{S-Multiple}

As already stated in the introduction, an appropriate description
of the environment within which the particle acceleration takes
place involves considering the coexistence of multiple
reconnecting current sheets. In this section, we present our
numerical results for the interaction of electrons and protons
with multiple current sheets taking parameter values
stochastically from a homogeneous distribution, together with a
theoretical analysis of these results.

\subsection{Numerical results}
\begin{figure}    %%%%%%%%%%%%%%%%%% FIGURE 5
   \centerline{\includegraphics[width=1.\textwidth,clip=]{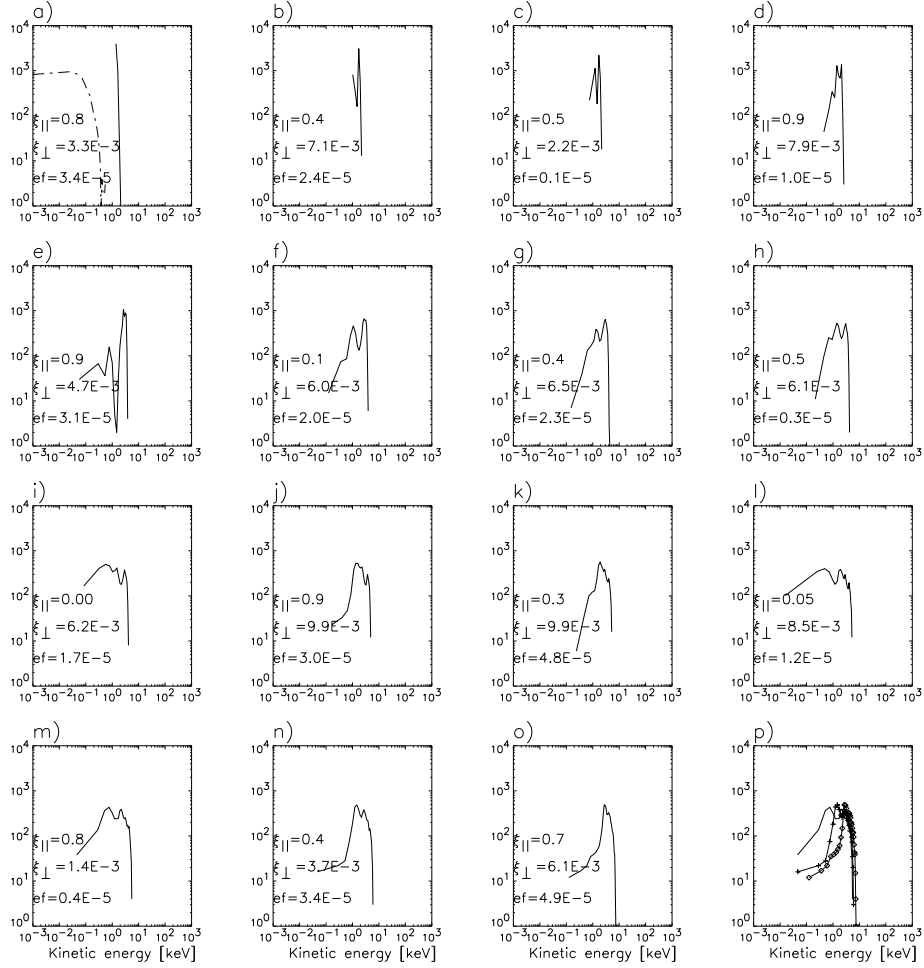}
              }
              \caption{Consecutive acceleration of 5000 electrons, initially with thermal
              velocities from 15 X-points. Each panel from left to right and from top
              to bottom shows the final kinetic energy distribution of the electrons after the
              interaction with each X-point. In the first panel, the dashed curve is the initial
              thermal kinetic energy distribution.
              Panel (p) shows the three last distributions on the same panel.}
   \label{F-Scatter-elec}
   \end{figure}
We performed numerical simulations of particles interacting
subsequently with 15 current sheets. In a previous work
(Gontikakis, Anastasiadis and Efthymiopoulos 2007) we studied the
cases of electrons and protons accelerated successively from
Harris-type current sheets. In the present work we present the
acceleration of protons and electrons from X-point reconnecting
current sheets. We consider 5000 particles that have initially a
thermal kinetic energy at $2 \times 10^6$~K. They are injected
inside the first X-point of the series. The injection points of
each particle is at $y=\pm 0.9$ and their initial position along
the x-axis varies randomly inside the range $-0.5<x<0.5$. The
particles with modified kinetic energies after the interaction
with the first X-point are injected through the next RCS. This
procedure is repeated 15 times. The parameters $\xi_\parallel$ and
$\xi_\perp$ for each RCS are selected randomly in the range:
$10^{-4}<\xi_\perp<10^{-2}$, $0.01<\xi_\parallel<1$. Similarly the
normalized electric field is randomly selected in the range
$10^{-6}<\epsilon<5\times 10^{-5}$ for electrons and  $1.84\times
10^{-3}<\epsilon<9.2\times 10^{-2}$ for protons. The orientation
of the accelerated particles velocities is randomized every time
before their injection into the next current sheet.

\begin{figure}    %%%%%%%%%%%%%%%%%% FIGURE 6
   \centerline{\includegraphics[width=1.\textwidth,clip=]{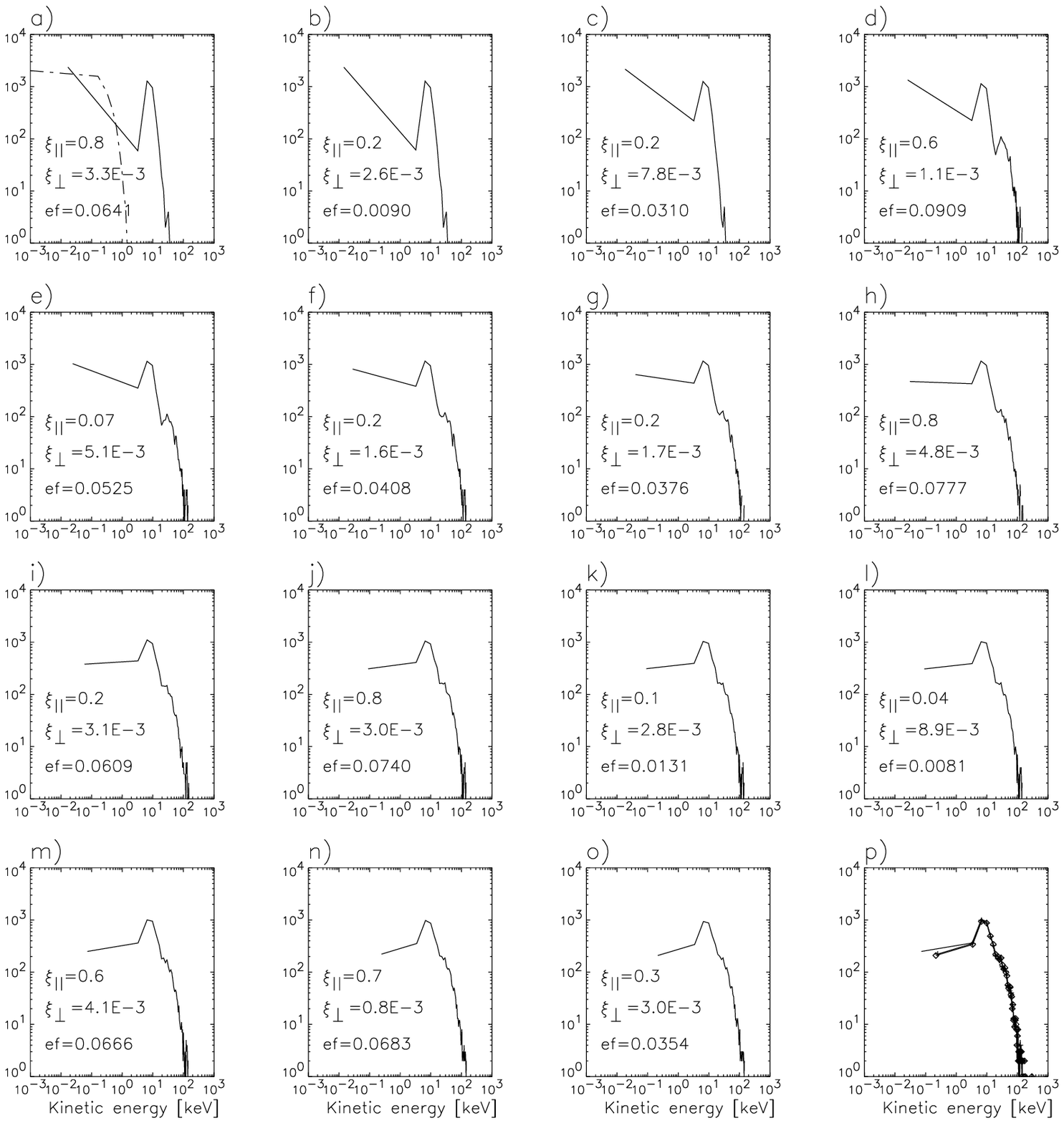}
              }
              \caption{Same as for Figure~\ref{F-Scatter-elec} for protons.}
   \label{F-Scatter-pro}
   \end{figure}

The resulting distributions are presented in
Figure~\ref{F-Scatter-elec} for electrons and in
Figure~\ref{F-Scatter-pro} for protons. The main result found
again is that the evolution of the kinetic energy distribution
converges towards a final state after a small number of current
sheets encounters. This happens for both types of particles and in
both types of current sheet configurations. In
Figures.~\ref{F-Scatter-elec}, \ref{F-Scatter-pro} in panel (p)
one can compare the last three kinetic energy distributions (13th,
14th and 15th) and conclude that have the same shape at high
energies. The same result was also found for the case of multiple
encounters with Harris-type current sheets (see Gontikakis,
Anastasiadis and Efthymiopoulos 2007).

\begin{figure}    %%%%%%%%%%%%%%%%%% FIGURE 7
   \centerline{\includegraphics[width=1.1\textwidth,clip=]{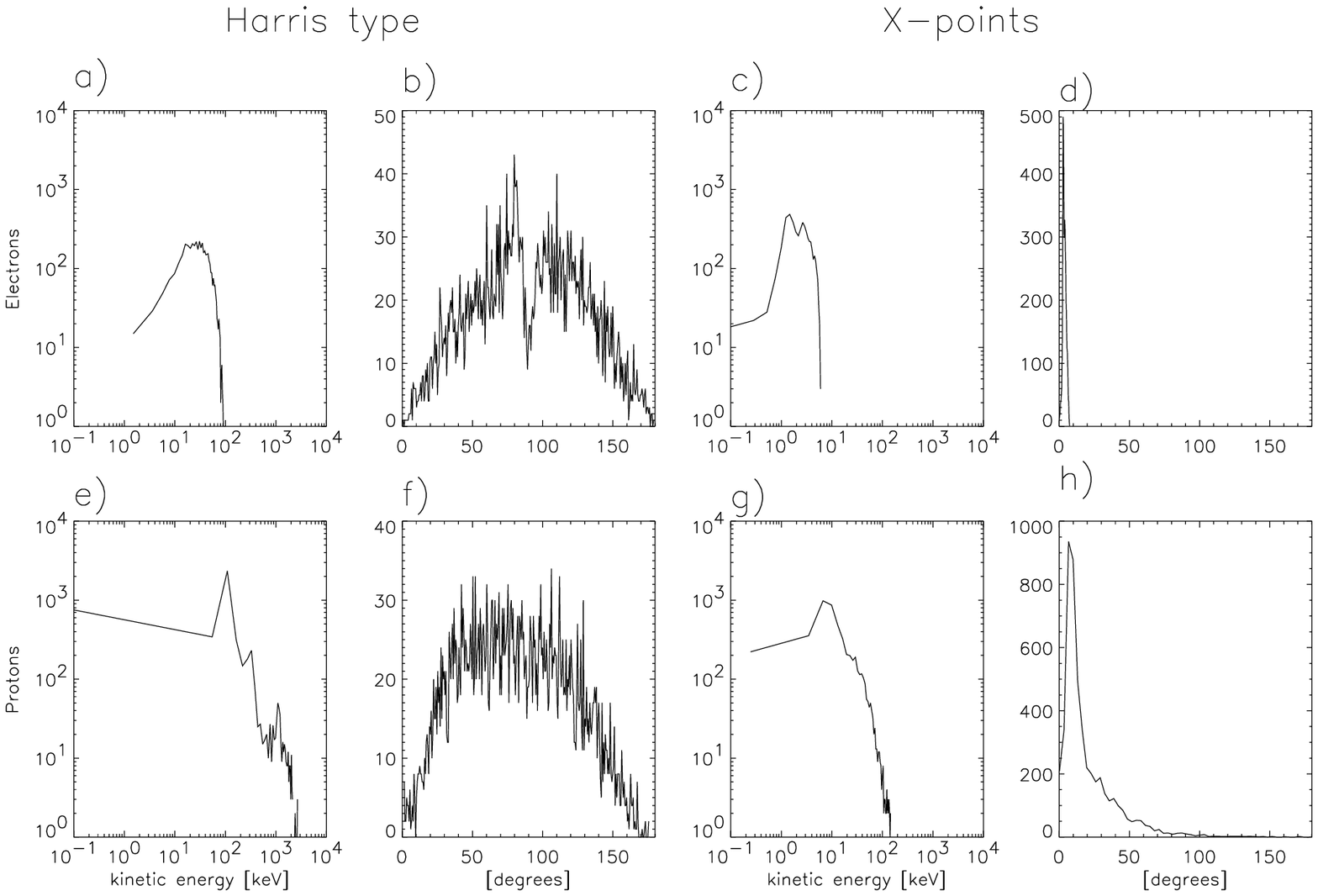}
              }
              \caption{Distributions of kinetic energy and pitch angles of multiple encounters
              of protons and electrons with Harris-type (panels a,b, e and f) and X-point
              (panels c, d, g and h) current sheets.
              The first row is for electrons and the second
              row is for protons. Panels a,c,e,g shows the final kinetic energy of the particles
              whereas panels b,d,f,h the final pitch angle distributions.}
   \label{F-distributions-multi}
   \end{figure}

For the case of X-points, the maximum kinetic energy gain is of
the order of 10~keV for electrons and  of 100~keV for protons. In
Figure~\ref{F-distributions-multi} one can compare the final
kinetic energy distributions for multiple encounters with
Harris-type (panels a,e) with the X-points (panels c,g). As was
the case for single encounters, particles through Harris-type
reconnecting current sheets, gain more energy by a factor of 10 in
respect to energy gain through their interactions with X-point
current sheet configuration.

\subsection{Theoretical analysis}

In Gontikakis, Anastasiadis and Efthymiopoulos (2007) a heuristic
explanation was given for the asymptotic convergence of the
kinetic energy distribution in a multiple particle- RCS
interaction. The explanation was that as the input kinetic energy
increases, the probability of an RCS, with random parameters
$(\xi_\perp,\xi_\parallel, \epsilon)$ within a specified parameter
space, to be an efficient accelerator decreases. We hereby provide
a detailed theoretical treatment of these results.

\begin{figure}    %%%%%%%%%%%%%%%%%% FIGURE 8
   \centerline{\includegraphics[width=1.0\textwidth,clip=]{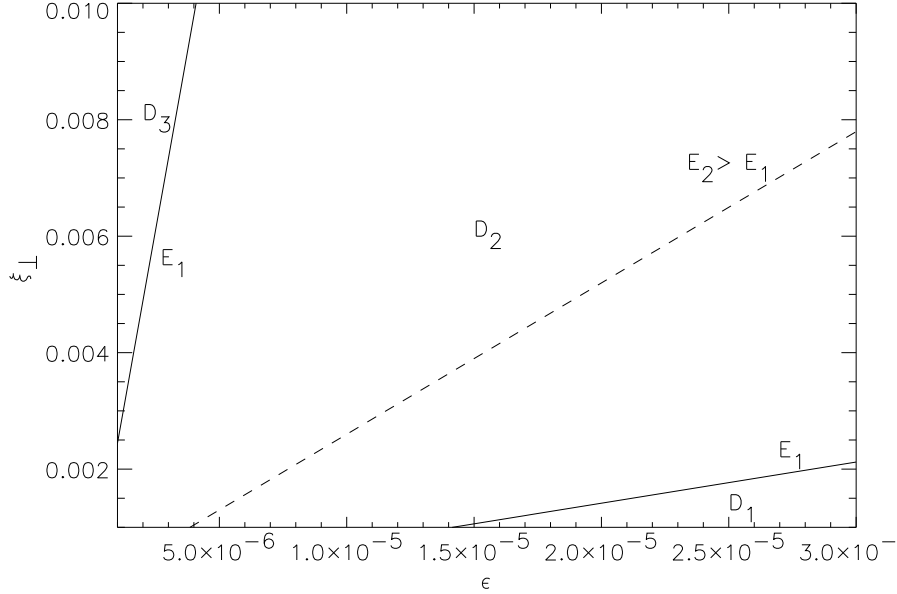}
              }
              \caption{The square box $10^{-6}\leq\epsilon\leq 3
              \times 10^{-5}$, $10^{-3}\leq \xi_\perp \leq
              10^{-2}$. The lower and upper solid lines are the
              graphs of Eqs.~(\ref{wclim}) and (\ref{xilim}) respectively
              for $w_0=1$, $E_0=E_1=4.7$~keV. A kinetic energy gain
              $w_0\,E_1$ is possible for all values of
              $\xi_\parallel$ in the domain $D_1$ and for some
              values of $\xi_\parallel$ in the domain $D_2$, while
              it is not possible in the domain $D_3$. The dashed
              line is the graph of Equation~(\ref{xilim}) for $w_0=1$
              and $E_0=E_2=47$~keV.}
   \label{F-domains}
   \end{figure}

Let $w_0$ be the maximum percentage of kinetic energy gain for
particles with initial energy $E_0$ interacting with a Harris-type
RCS with parameters $(\xi_\perp,\xi_\parallel,\epsilon)$. We shall
consider in detail the case of particles entering the RCS at $z=0$
with $\dot{x}_0=0$. Thus $I_2=-\xi_\parallel y_0$. Setting
$E_{max}=w_0E_0$ in Equation~(\ref{emax}) yields:
\begin{equation}\label{xipareq}
\xi_\perp\xi_\parallel\Delta y +\epsilon +
\sqrt{2\epsilon\xi_\perp\xi_\parallel\Delta y +\epsilon^2
+2\xi_\perp^2E_0} = {w_0\xi_\perp^2E_0\over 2}~~,
\end{equation}
where $\Delta y = y_{out}-y_0$. Equation~(\ref{xipareq}) can be used in
order to find the critical value of $\xi_\parallel$ above which
the kinetic energy gain surpasses the percentage $w_0$ for fixed
values of the remaining parameters. Setting $\xi_\parallel=0$ in
Equation~(\ref{xipareq}) yields the minimum percentage:
\begin{equation}\label{wm}
w_m={\epsilon\over\xi_\perp^2E_0}\big(\epsilon
+\sqrt{\epsilon^2+2\xi_\perp^2E_0}\big)~~.
\end{equation}
If $w_0>w_m$ Equation~(\ref{xipareq}) admits a positive solution for
$\xi_\parallel$ provided that:
\begin{equation}
\xi_\parallel\leq {w\xi_\perp E_0\over\epsilon\Delta y}
-{\epsilon\over\xi_\perp\Delta y}~~.
\end{equation}
The solution reads:
\begin{equation}\label{xiparsol}
\xi_\parallel = {1\over\Delta y}\bigg[{w_0\xi_\perp
E_0\over\epsilon} -\sqrt{2(w_0+1)E_0}\bigg]~~.
\end{equation}
If $\xi_\parallel$ is greater than the value (\ref{xiparsol}), the
percentage of kinetic energy gain is larger than $w_0$. We now
wish to estimate, for fixed $(w_0,E_0)$ the probability within the
parameter space that an RCS will accelerate the particles by a
percentage larger or equal than $w_0$. This can be done with the
help of Figure~\ref{F-domains}. The square box shows the domain of
the parameter space for $\epsilon$ and $\xi_\perp$, which in our
simulations is $10^{-6}\leq\epsilon\leq 3\times 10^{-5}$,
$10^{-3}\leq\xi_\perp\leq 10^{-2}$. The lower solid line
corresponds to the equation:
\begin{equation}\label{wclim}
\xi_\perp={\epsilon\over w_0}\sqrt{2(w_0+1)\over E_0}~~.
\end{equation}
This line yields the locus of values of $(\epsilon,\xi_\perp)$ for
which the given percentage $w_0$ coincides with the minimum
possible percentage of kinetic energy gain for the given energy
$E_0$, according to Equation~(\ref{wm}). This locus divides the square
into two domains. In the lower domain ($D_1$) one has $w_m>w_0$,
thus for all values of $\xi_\parallel$ the kinetic energy gain is
higher than $w_0$. On the other hand, in the upper domains ($D_2$
and $D_3$) one has $w_m<w_0$, thus only for $\xi_\parallel$ large
enough the percentage $w_0$ can be surpassed. Now, the upper solid
line yields the locus at which the minimum value of
$\xi_\parallel$ for which the percentage $w_0$ is reached exceeds
the upper allowed value $\xi_{\parallel,max}$ within the selected
parameter space (in our examples $\xi_{\parallel,max}=1$). This is
given by
\begin{equation}\label{xilim}
\xi_\perp={\epsilon\over w_0}\bigg(\sqrt{2(w_0+1)\over E_0}
+{\Delta y\over w_0E_0}\bigg)~~.
\end{equation}
In the domain $D_3$ above the line (\ref{xilim}) we have
$\xi_\parallel>\xi_{\parallel,max}$, thus this domain lies
entirely outside the parameter space. In view of the above
analysis, the total subvolume of the parameter space in which the
kinetic energy gain exceeds the percentage $w_0$ is given by
\begin{equation}\label{subvol}
V(w_0,E_0)=(\xi_{\parallel,max}-\xi_{\parallel,min})S(D_1)+
\int_{D_2}{\big(\xi_{\parallel,max}-\xi_\parallel(w_0,E_0,\xi_\perp,\epsilon)\big)}
d\xi_\perp d\epsilon
\end{equation}
where $\xi_{\parallel,max}$, $\xi_{\parallel,min}$ are the limits
of $\xi_\parallel$ in the selected parameter space, $S(D_1)$ is
the surface of the domain $D_1$ and
$\xi_\parallel(w_0,E_0,\xi_\perp,\epsilon)$ is the solution
(\ref{xiparsol}) for $\xi_\parallel$. The probability of an RCS to
accelerate the particles by a gain factor larger than $w_0$ is
given by:
\begin{equation}\label{prob}
P(w_0,E_0)={V(w_0,E_0)\over V_{total}}~~.
\end{equation}
This probability is roughly proportional to the total area of the
domains $D_1$ and $D_2$ over the total area of the square box. We
can see now that as $E_0$ increases, the probability $P(w_0,E_0)$
decreases for any fixed value of $w_0$. This is simply because,
according to Eqs.(\ref{wclim}) and (\ref{xilim}) the slopes of
both limiting lines decrease, thus the total area $S(D_1)+S(D_2)$
also decreases. In fact, beyond a sufficiently high value of
$E_0$, at which the upper limiting line touches the lower right
apex of the square, the domains $D_1$ and $D_2$ both shrink to a
null domain. Thus the probability of a gain factor $w_0$ shrinks
to zero. This was precisely found numerically in Figure~9 of
Gontikakis, Anastasiadis and Efthymiopoulos (2007) (note an error
in this figure: $w$ there corresponds to $w_0-1$ in our notation).
Setting the value $w_0=1$ (corresponding to $w=0$ in Gontikakis,
Anastasiadis and Efthymiopoulos 2007), as well as
$\xi_\perp=10^{-3}$ and $\epsilon=3\times 10^{-5}$ for the lower
right appex of the square, and $\xi_{\parallel,max}=1$, we find a
maximum energy through Equation~(\ref{xipareq}) equal to
$E_0\simeq 7.5\times 10^{-2}$, or, in physical units
$E_0=4.2\times 10^2$ keV, which agrees well with the limit found
numerically in Gontikakis, Anastasiadis and Efthymiopoulos (2007).

\section{Conclusions}
\label{S-Summary}

In the present work we studied the acceleration of electrons and
protons in a multiple particle-RCS interaction scenario, assuming
two different reconnection topologies, the Harris-type current
sheet and the X-point. Our main conclusions are the following:

1) The particles accelerated close to a X-point, with specific
field parameters ($\xi_\perp, \xi_\parallel, \epsilon$) gain a
smaller amount of kinetic energy than particles accelerated
through a Harris-type current sheet with the same model
parameters. This is because particles accelerated through an
X-point encounter a stronger average perpendicular magnetic field
component.

2) The particles escape from a Harris-type current sheet in longer
times than from an X-point.

3) A previously derived analytical formula (Equation~\ref{emax}),
yielding the maximum kinetic energy gain for particles accelerated
through a Harris type topology (Gontikakis, Efthymiopoulos and
Anastasiadis 2006), provides a satisfactory approximation to the
kinetic energy gains computed numerically. (Equation~\ref{emax})
can also be modified to yield an estimate for the kinetic energy
gain of particles accelerated by an X-point if the perpendicular
component $\xi_\perp$ is replaced by an effective field
$<\xi_\perp>$ which is the mean perpendicular magnetic field along
an orbit.

4) A number of similarities are also found for the acceleration of
particles through Harris-type RCSs and X-points. An important
fraction of protons with initially thermal distributions are not
accelerated at all. The pitch angle distributions of the
accelerated electrons and protons, present sharp peaks for angles
$<10$\degr.

5) When the particles encounter multiple RCSs with parameters
obtain randomly from a uniform sample, the particles kinetic
energy distribution tends to acquire a limiting form after some
iterations. This tendency was previously observed (Gontikakis,
Anastasiadis and Efthymiopoulos 2007) in the case of multiple
encounters with Harris-type RCSs, and we here confirm it also in
the case of X-points. A theoretical explanation is given by
considering the probability of an RCS to act as efficient
accelerator (parameterized by the percentage $w_0$ of kinetic
energy gain with respect to some initial energy $E_0$). In
particular we demonstrate analytically that this probability
decreases as $E_0$ increases.

%%%%%%%%%%%%%%%%%%%%%%%%%%%%%%%%%%%%%%%%%%%%%%%%%%%%%%%%%%%%%%%%%%%%%%%%%%%
\begin{acks}
We wish to thank Dr. I. Contopoulos for stimulating discussions on
the problem of magnetic reconnection as well as the anonymous
referee and Dr. K-L. Klein for their comments and suggestions that
improved our manuscript. The work of CG and CE was supported in
part by the Research Committee of the Academy of Athens. AA would
like to thank the members of the Particle Acceleration Working
Group of the CESRA-2007 workshop for many stimulating discussions.
\end{acks}

\end{article}
\end{document}